\documentclass{rspublic}

\usepackage{graphicx}
\usepackage{epic,eepic}
\usepackage{graphics}
\usepackage{epsfig}

\usepackage{subfigure}

\renewcommand{\pi}{{\partial_i}}

\newcommand{\be}{\begin{equation}}
\newcommand{\ee}{\end{equation}}
\newcommand{\bea}{\begin{eqnarray}}
\newcommand{\eea}{\end{eqnarray}}

\usepackage{amsmath}
\usepackage{amsfonts}
\usepackage{amssymb}
\usepackage{bm}
\usepackage{color}
\usepackage{graphicx}

\begin{document}
\title[Shear-Banding from lattice kinetic models]{Shear Banding from lattice kinetic models with competing interactions}
\author[Benzi et al.]{Roberto Benzi$^{1}$ , Mauro Sbragaglia$^{1}$ , Massimo Bernaschi$^{2}$  \& Sauro Succi$^{2}$ }
\affiliation{$^{1}$ Department of Physics and INFN,University of Tor Vergata,  Via della Ricerca Scientifica 1, 00133 Rome, Italy\\
$^{2}$ Istituto per le Applicazioni del Calcolo CNR, Viale del Policlinico 137, 00161 Roma, Italy}
\maketitle

\begin{abstract}{Soft Glassy Materials, Non Linear Rheology, Lattice Kinetic models, frustrated phase separation}
We present numerical simulations based on a Boltzmann kinetic model 
with competing interactions, aimed at characterizating the rheological 
properties of soft-glassy materials. 
The lattice kinetic model is shown to reproduce typical signatures of
driven soft-glassy flows in confined geometries, such as 
Herschel-Bulkley rheology, shear-banding and histeresys. 
This lends further credit to the present lattice kinetic model as a 
valuable tool for the theoretical/computational investigation of the
rheology of driven soft-glassy materials under confinement.
\end{abstract}

\maketitle

\section{Introduction}

Many applications in modern science, engineering and biology have prompted 
the recent blossoming of research in the rheology of soft-flowing and nonergodic materials, such as emulsions, foams, gels  (Larson 1999,Coussot 2005, Chaikin \& Lubensky 1995, Lyklema 1991, Evans \& Wennerstrm 1999, Degennes 1979, Doi \& Edwards 1986, Grosberg \& Khokhlov 1994, Weaire \& Hutzler 1999). The theoretical understanding of such materials raises interesting questions on its own. Indeed, since soft-materials share simultaneously many distinctive features of the three basic states of matter (solid, liquid and gas), their quantitative description does not fall within the traditional methods of equilibrium and/or non-equilbrium statistical mechanics. New concepts and theoretical paradigms are required to adjust many non-standard features, such as rheology, long-time relaxation, dynamic disorder, ageing and related phenomena. In particular,  under simple shear conditions, some of these complex fluids may separate into bands of widely different viscosities. 
This phenomenon, known as ``shear banding'' (Berret 2005), involves inhomogeneous flows
where macroscopic bands with different shear rates or shear stresses coexist in the sample. 
Although shear banding is attributed to a shear-induced transition from a microscopic organization of the fluid structure to another, it still raises lots of theoretical and experimental challenges (Manneville {\it et al.} 2007, Sollich {\it et al.} 2009, Fardin {\it et al.} 2010).\\
Simultaneously, the need for new tools of analysis also emerges for computational studies, which typically involve many interacting space and time scales, the latter case being particularly acute in view of  the aforementioned long-time relaxation.  In the recent past, we presented a new conceptual/computational scheme for the numerical simulation of soft-flowing materials (Benzi {\it et al.} 2009, Bernaschi {\it et al.} 2009, Benzi {\it et al.} 2010). The scheme is based on a (Lattice) Boltzmann (LB) formulation (Bathnagar, Gross  \& Krook 1954, Benzi {\it et al.} 1992, Chen \& Doolen 1998, Gladrow 2000)  for interacting binary fluids (Shan \& Chen 1993,1994), in which, by a proper combination of short-range attraction  and mid-range repulsion (competing self interactions), an effective form of frustration was encoded within the physics of the binary mixture (Benzi {\it et al.} 2009). More specifically, by tuning the above interactions in such a way as to bring the surface tension down to nearly vanishing values, many typical signatures of soft-glassy behaviour, such as long-time relaxation, dynamical arrest, ageing, and non-linear Herschel-Bulkley rheology, have been clearly detected in numerical studies (Benzi {\it et al.} 2009, Bernaschi {\it et al.} 2009, Benzi {\it et al.} 2010). 
In this paper, we further elaborate along these lines by performing numerical 
simulations of confined flows and looking at their rheological properties, i.e. the relation between the applied shear and the developed stress by the fluid. 
By performing numerical simulations over a wide range of shear rates, we detect 
the emergence of shear banding effects, i.e. a stress plateau, which is shown 
to correspond to a wide range of stationary shear rates.  
These results lend further credit to the Boltzmann kinetic formulation as a valuable tool 
for the systematic study of the emergence of non linear rheological properties 
from mesoscopic interactions.
\section{The Model: Lattice Boltzmann equation with competing self interactions}

\begin{figure}[!t]
\begin{center}
\includegraphics[scale=0.3]{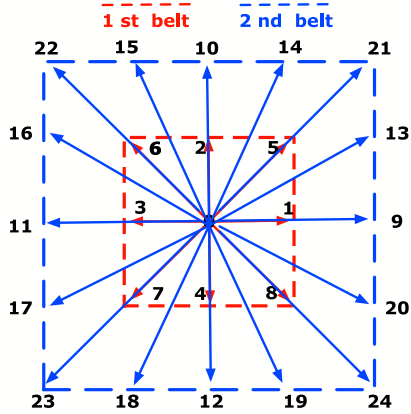}
\caption{The discrete $25$-speed lattice. Both belts are illustrated with the corresponding velocities. Further details can be found in the paper by Benzi and coworkers (Benzi {\it et al.} 2009).}
\label{fig:2}
\end{center}
\end{figure}

In this section we review the main properties of the lattice Boltzmann model used in the numerical simulations. Further details can be found in our recent work (Benzi {\it et al}. 2009). The starting point is the lattice transcription of the generalized Boltzmann equation (Bathnagar, Gross  \& Krook 1954, Benzi {\it et al.} 1992, Chen \& Doolen 1998, Gladrow 2000) for a multicomponent fluid with $S$ species, as inspired by the work of Shan \& Chen (Shan \& Chen 1993, Shan \& Doolen 1995):
\begin{equation}
f_{is}(\vec{r} + \vec{c}_i \Delta t, t + \Delta t)-f_{is}(\vec{r} , t ) =  -\frac{\Delta t}{\tau_s}[f_{is}(\vec{r} , t )-f_{is}^{(eq)}(\rho_s ,  \vec{u}+\tau_s \vec{F}_{s} /\rho_s)].
\label{eq:be}
\end{equation}
In the above, $f_{is}(\vec{r},t)$ is the probability density function of finding a particle of species  $s=1...S$ at site $\vec{r}$ and time $t$, moving along the $i$-th lattice direction, defined  by the discrete speeds $\vec{c}_i$ with $i=0...b$ (see Figure 1).  
For simplicity, the characteristic propagation time lapse $\Delta t$ is taken  equal to unity ($\Delta t =1$) in the following.   The left hand-side of (\ref{eq:be}) stands for molecular free-streaming, whereas the  right-hand side represents collisions as a simple time relaxation towards the local  Maxwellian equilibrium $f_{is}^{(eq)}(\rho_s ,\vec{u} )$ on a time scale $\tau_s$.  The local Maxwellian is truncated at second order, an approximation which suffices
to recover the correct athermal hydrodynamic balance 
$$
f_{is}^{(eq)}(\rho_s,\vec{u} )=
w_i^{(eq)} \rho_s \left(1+\frac{({u}_a c_{ia})}{c_S^2}+ \frac{({c}_{ia} {c}_{ib}-c_S^2 {\delta}_{ab})}{2 c_S^4}{u}_a {u}_b \right)
$$
with $c_S^2$ the square of the sound speed in the model and  ${\delta}_{ab}$ the Kronecker delta with $a,b$ indicating the Cartesian components  (repeated indices are summed upon).   The $w_i^{(eq)}$'s are equilibrium weights used to enforce isotropy of the hydrodynamic  equations (Benzi {\it et al.} 1992, Gladrow 2000, Chen \& Doolen 1998).  The equilibrium for the $s$ species is a function of the local species density
$$\rho_s(\vec{r}, t)=\sum_i f_{is}(\vec{r}, t)$$
and the common velocity, defined as
$$\vec{u} (\vec{r}, t)= \frac{\sum_s   \sum_i f_{is}(\vec{r}, t) \vec{c}_i }{ \sum_s  \rho_s (\vec{r}, t)}.$$
The common velocity receives a shift from the force $\vec{F}_{s}$ acting on the $s$ species (Shan \& Chen 1993, Shan \& Doolen 1995).   This force may be an external one or due to intermolecular interactions.   
The pseudo-potential forces embed the essential features to achieve 
phase separation (non-ideal equation of state and non-zero surface tension), as 
well as a mechanism of frustration through competing self-interactions.  More specifically, within each species, the forces consist of an attractive (denoted with $(att)$) component, acting  only on the first Brillouin region (belt, for simplicity), and a repulsive (denoted with $(rep)$) one acting  on both belts, whereas  the force between species $(X)$ is short-ranged and repulsive. In equations:
$$
\vec{F}_s(\vec{r}, t) = \vec{F}^{(att)}_s(\vec{r}, t) + \vec{F}^{(rep)}_s(\vec{r}, t)+\vec{F}^X_s(\vec{r},t)
$$
where
\be
\vec{F}^{(att)}_s(\vec{r}, t) = -G^{(att)}_s \psi_s(\vec{r},t) \sum_{i \in belt 1} w_i \psi_s(\vec{r}_{i},t) \vec{c}_{i}  
\ee
\be
\vec{F}^{(rep)}_s(\vec{r}, t) =  - G^{(rep)}_{s} \psi_s(\vec{r},t) \sum_{i \in belt 1} p_{i} \psi_s(\vec{r}_{i},t) \vec{c}_{i}  - G^{(rep)}_{s} \psi_s(\vec{r},t) \sum_{i \in belt 2} p_{i} \psi_s(\vec{r}_{i},t) \vec{c}_{i}  
\ee
\be\label{eq:last}
\vec{F}_s^X (\vec{r},t) = - \frac{1}{(\rho_0^{(s)})^2}\rho_s (\vec{r},t) \sum_{s' \neq s}\sum_{i \in belt 1}   G_{s s'}w_i \rho_{s'}(\vec{r}_i,t) \vec{c}_i.
\ee
In the above, the groups ``belt $1$'' and ``belt $2$''  refer to the first  and second Brillouin zones in the lattice and $\vec{c}_{i}$, $p_{i}, w_i$ are  the  corresponding discrete speeds and associated weights (see figure \ref{fig:2} and table \ref{T1}). The interaction parameter $G_{ss'}=G_{s's}$, $s' \ne s$, is the cross-coupling between species, $\rho_0$ is a reference  density to be defined shortly and, finally, $\vec{r}_{i} = \vec{r}+\vec{c}_{i}$  are the displacements along the $i$-th directions. These interactions are sketched in Figure \ref{fig:2} for the case of a two component fluid (say species A and B).
This  model bears similarities to  the next-to-nearest-neighbor frustrated lattice  spin models (Shore \& Sethna 1991, Shore, Holzer \& Sethna 1992). 
However, in our case, a high-lattice connectivity (Sbragaglia {\it et al.} 2007) is required  to ensure compliance with macroscopic non-ideal hydrodynamics, which is at the core of the complex rheology to be discussed in this work.  To this purpose, the first belt is discretized with $8$ speeds, whereas the  second  with $16$, for a total of $b=25$ connections (including rest particles). All the weights take the values illustrated in Table \ref{T1}.  

\begin{table}
\begin{center}
\begin{tabular}{l l}
\hline
 \qquad \qquad \qquad Forcing Weigths (for $\vec{F}^{(rep)}_s$) \\
\hline
  $p_{i}  = 247/420\;     $&$ \quad i = 0$\\
  $p_{i}  = 4/63\;     $&$ \quad i = 1,4$\\
  $p_{i}  = 4/135 \;   $&$ \quad i = 5,8$\\
  $p_{i}  = 1/180 \;   $&$ \quad i = 9,12$\\
  $p_{i}  = 2/945 \;   $&$ \quad i = 13,20$\\
  $p_{i}  = 1/15120 \; $&$ \quad i = 21,24$\\
\hline
\end{tabular}

\vspace{.2in}

\begin{tabular}{l l}
\hline
 \qquad \qquad \qquad Forcing Weights (for $\vec{F}^{(att)}_s$ and $\vec{F}^X_s$) \\
\hline
  $w_{i}  = 4/9  $&$\quad i = 0$\\
  $w_{i}  = 1/9  $&$\quad i = 1,4$\\
  $w_{i}  = 1/36 $&$\quad i = 5,8$\\
\hline
\end{tabular}

\end{center}
\caption{Links and weights of the two belts, $25$-speeds lattice. \label{T1}}
\end{table}

The present model has already proven capable of reproducing several signatures
of soft-glassy behaviour, such as long-time relaxation, ageing, and non-Newtonian
rheology. As a general remark, the onset of non-trivial behaviour has been found
to associate with the regime of very-low surface tension, a physical quantity
that the present scheme allows to tune through the coupling parameters $G$'s and
$\rho_0$. In particular, once the couplings $G$'s are fixed, surface tension can be brought down to nearly vanishing values by increasing the value of $\rho_0$, because the repulsive intra-species force, $\vec{F_s}^X$, contributing a positive surface tension, are scaled by a factor $1/\rho_0^2$, see eq. (\ref{eq:last}). Full details can be found in our previous publications. 
To date, the aforementioned effects have been explored only in  homogeneous, boundary-free geometries. Given the importance of relating to experimental results, it is of great interest to assess whether the above phenomenology can also be reproduced for the case of soft-glassy flows under geometrical confinement. This is precisely the task undertaken in the following section.

\section{Numerical Results}
We have simulated a two-dimensional channel flow  driven by the upper wall, moving at speed $U_w$ along  the  mainstream direction, $x$.  The channel measures $L=256$ lattice sites in length and $H=512$  lattice sites in width. The initial conditions correspond to a modulated density,  $\rho_A(x,y, t=0)=\rho_m \left(1+0.1 \sin \left( 6.28 \frac{x}{16} \right) \sin \left(6.28 \frac{y}{16} \right) \right)$, $\rho_B(x,y, t=0)=\rho_m \left(1+0.01 \sin \left( 6.28 \frac{x}{16} \right) \sin \left(6.28 \frac{y}{16} \right) \right)$ with $\rho_m=0.612$ and zero flow. Periodic boundaries are imposed at the inlet and outlet sections. The coupling parameters are as follows $G_A^{(att)}=-9.0$, $G_A^{(rep)}=8.1$, $G_B^{(att)}=-8.0$, $G_B^{(rep)}=7.1$, and $G_{AB}=0.405$. The time relaxation is $\tau=1$ for both species, corresponding to a kinematic viscosity $\nu=1/6$ in lattice units. Under ordinary (Newtonian) flow conditions, the wall drive generates  a linear flow profile $u_x(y) = U_w y/H$, with  an associated homogeneous stress $\sigma_{xy} = \mu U_w/H$, $\mu=\rho \nu$  being the dynamic viscosity of the fluid.    This is precisely the situation our model is expected to reproduce  in the regime of sufficiently high-surface tension.

\subsection{High surface tension: Newtonian rheology}

\begin{figure}[!t]
\begin{center}
\includegraphics[scale=0.6]{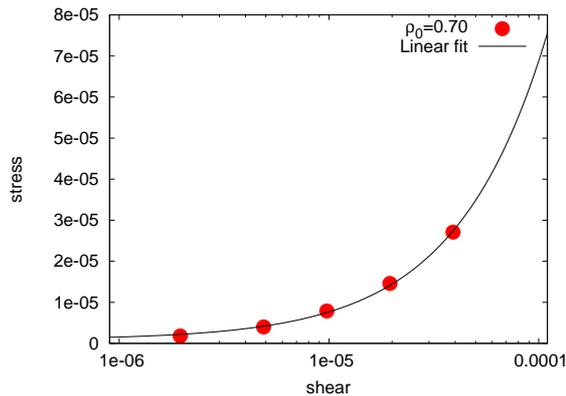}
\caption{Stress {\it vs} shear for the case $\rho_0=0.70$. The solid line is a linear fit (exponential in log representation).}
\label{Newton}
\end{center}
\end{figure}

In Figure \ref{Newton}, we report the average stress $\sigma_{xy}$, as  collected over the set of slices at $x=1,L$, as a function of the applied wall-shear $S_w = U_w/H$.   The simulations have been performed with $\rho_0=0.70$, corresponding to a sizeable value of surface tension ($\gamma \sim 0.05$), for which Newtonian behaviour is expected. As one can appreciate, the stress-strain relation, in log-scale, is  well fitted by an exponential curve, indicating a linear shear-stress relation.

\subsection{Low surface tension: non-newtonian rheology and shear banding}

In order to investigate the emergence of non-Newtonian behaviour, we have repeated the simulations with a higher $\rho_0=0.83$, corresponding to a  much lower surface tension $\gamma \sim 5 \times 10^{-3}$. The corresponding stress-shear relation is shown  in figure \ref{NNewton}. Several comments are in order. At low shear, $S_w < 2 \; 10^{-6}$, the stress-shear still follows a linear, newtonian relation, as in the high-surface tension case described previously.  Upon increasing the driving shear, however, no further increase of the stress is observed, up to $S_w= 2 \; 10^{-5}$.  This is the so-called {\it shear-banding} effect, signaling qualitative rearrangements in the structural  response of the flow pattern, to be discussed shortly. By further increasing the driving shear, the stress curve regains a linear (exponential in log-scale) behaviour, corresponding however to a  smaller, value of the effective viscosity. The emerging picture is that of a yield-stress, shear-thinning fluid, whose effective viscosity can be mapped into a Herschel-Bulkley relation of the form $\sigma_{xy} = A + B\;S_w^{\alpha}$, with $A=1.48 \times 10^{-6} $, $B=0.000441$ and $\alpha \sim 0.25$. The appearance of a non-zero yield stress, $\sigma_Y \equiv A$, prepares the ground for the emergence of shear banding. Indeed, the fluid can only flow in regions where the stress exceeds $\sigma_Y$ (the shear bands), while in the rest of the domain the fluid stands basically still, like a solid.

\subsection{Low surface tension: hysteresis}

\begin{figure}[!t]
\begin{center}
\includegraphics[scale=0.48]{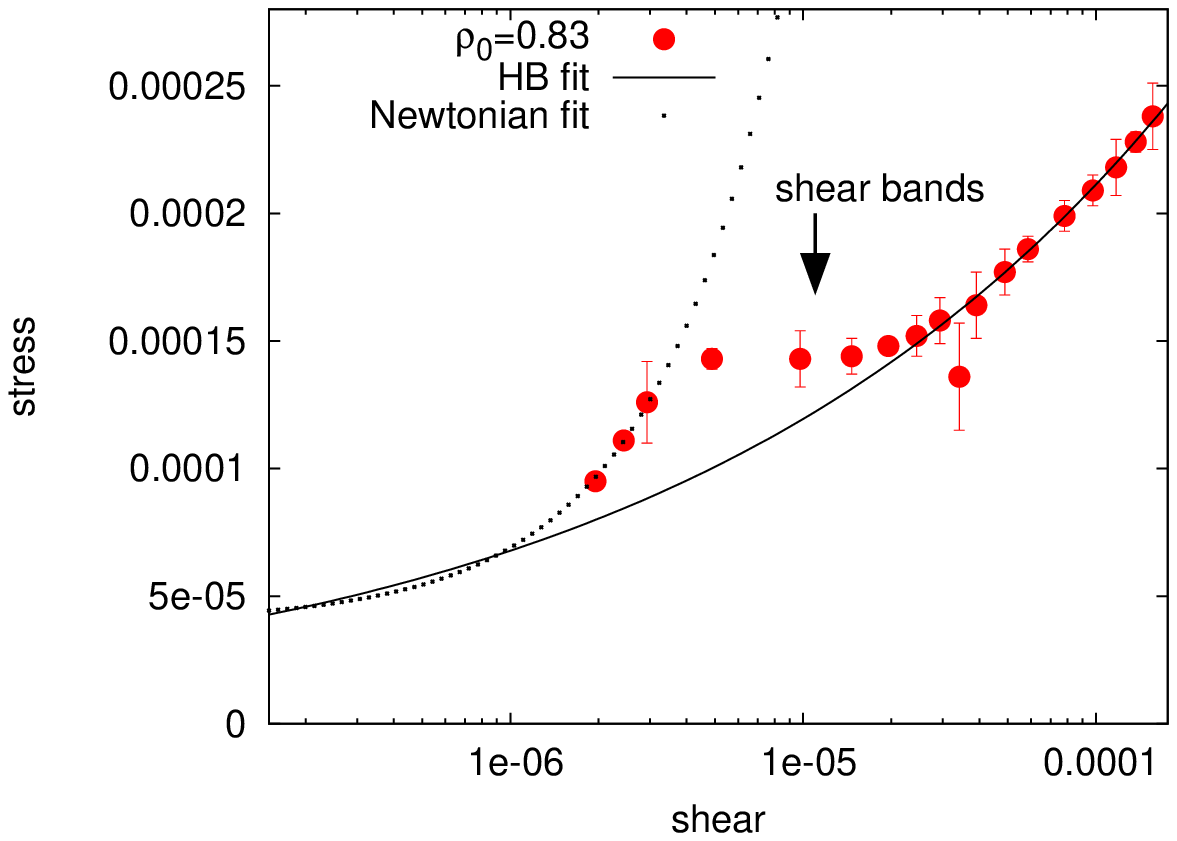}
\includegraphics[scale=0.48]{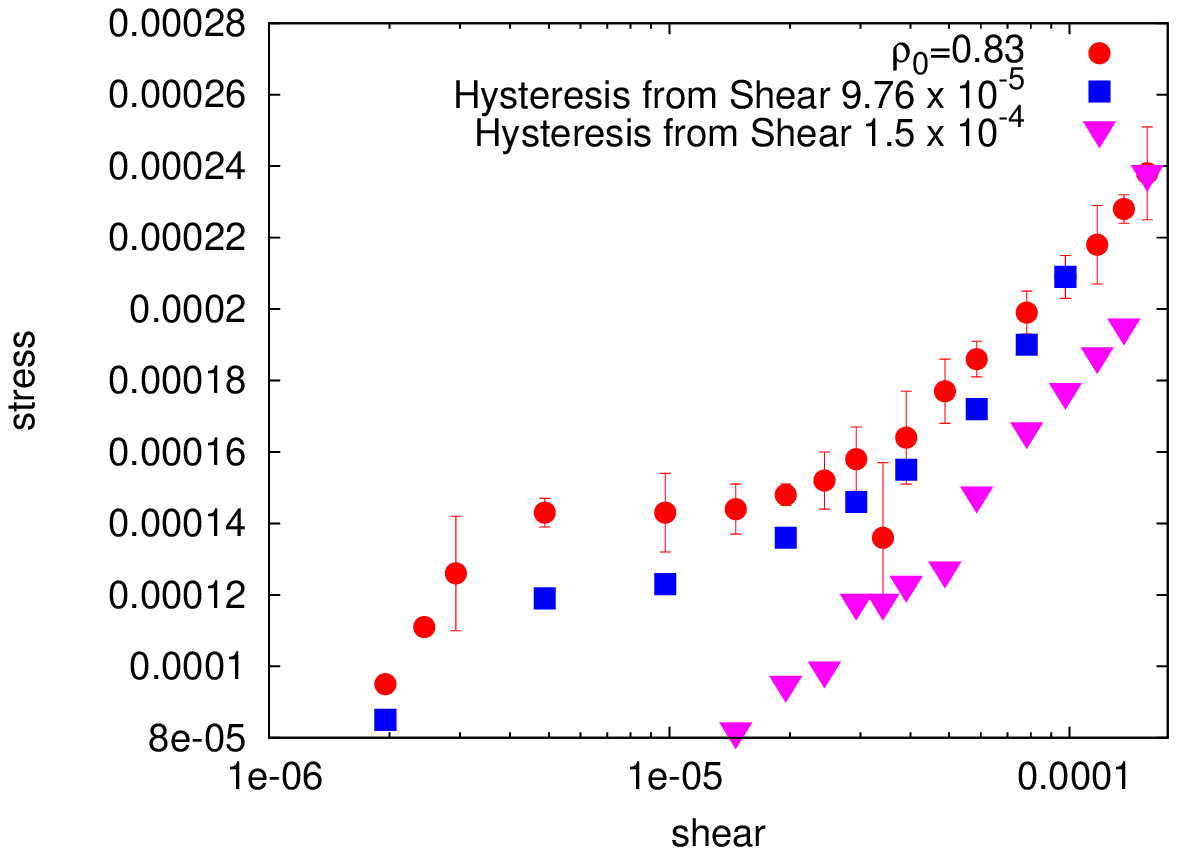}
\caption{Left: Stress {\it vs} shear for the case $\rho_0=0.83$. As we observe, there is a range of shear rates at which a Plateau in the stress tensor is observed. All the results have been obtained by averaging over many configurations and starting from the same initial condition. Right: we repeat the same measurements for the Stress {\it vs} shear starting from an initial configuration at high shear rates and decreasing the shear at the wall. Hysteresis effects are visible.
}
\label{NNewton}
\end{center}
\end{figure}

We have also investigated the occurrence of {\it memory-dependent phenomena}, such as hysteresis. To this purpose, we have repeated a series of simulations taking as an initial condition the steady state at $S_w=9.76 \times 10^{-5}$, and then scanning $S_w$ backwards to decreasing values. As again apparent from figure \ref{NNewton}, the corresponding values of the stress do {\it not} lie on the same curve as before, but trace in fact a  lower-lying curve.  A similar behaviour is observed by repeating the same backward scan, starting from a higher value $S_w=1.5 \times 10^{-4}$, although no crossover is observed in this latter case. By and large, all of the non-Newtonian features portrayed in this figure  reveal a striking similarity with recent experiments  on soft-glassy flows in microchannels (Manneville {\it et al.} 2007).

\subsection{Spatial velocity profiles}

\begin{figure}[!t]
\begin{center}
\includegraphics[scale=0.6]{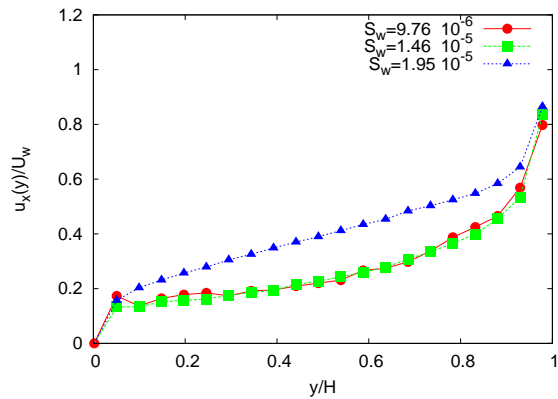}
\caption{Velocity profiles for $\rho_0=0.83$ and different wall velocities corresponding to the following shears $S_w=9.76 \times 10^{-6}, 1.46 \times 10^{-5}, 1.95 \times 10^{-5}$. The vertical coordinate has been normalized between $0$ and $1$ and the flow velocity has been normalized with respect to the corresponding wall velocity.}
\label{Profiles}
\end{center}
\end{figure}

Finally, we inspect the spatial distribution of  the average flow velocity, $u_x(y)$, inside the channel, see figure \ref{Profiles}. From this figure, it is apparent that in the low-surface tension, non-Newtonian regime, the velocity profiles tend to form a steep layer  next to the two walls (the shear bands), with a susbtantial flattening in the bulk region of the flow, where little dissipation takes place. This flattening is responsible for the plateau in the shear-stress relation. Much like a turbulent flow, under increasing shear, the non-newtonian fluid reacts in such a way as to confine most of the shear, and stress, to an increasingly thinner near-wall layer.   This is again in line with recent experimental observations (Manneville {\it et al.} 2007).

\section{Conclusions}
Summarizing, we have shown that the two-component
Lattice Boltzmann model with competing interactions is capable 
of reproducing distinctive non-Newtonian features of
driven soft-glassy flows in confined geometries, such
non-Newtonian rheology, shear-banding and hysteresis.
The simulation data bear striking similarities with 
recent experimental results of driven soft-glassy flows in
Couette geometries. Future work shall be directed to the quantitative
comparison with experimental data, as well as to the investigation
of the implications of low surface tension on the global and local
rheology of soft-glassy materials.  


\end{document}